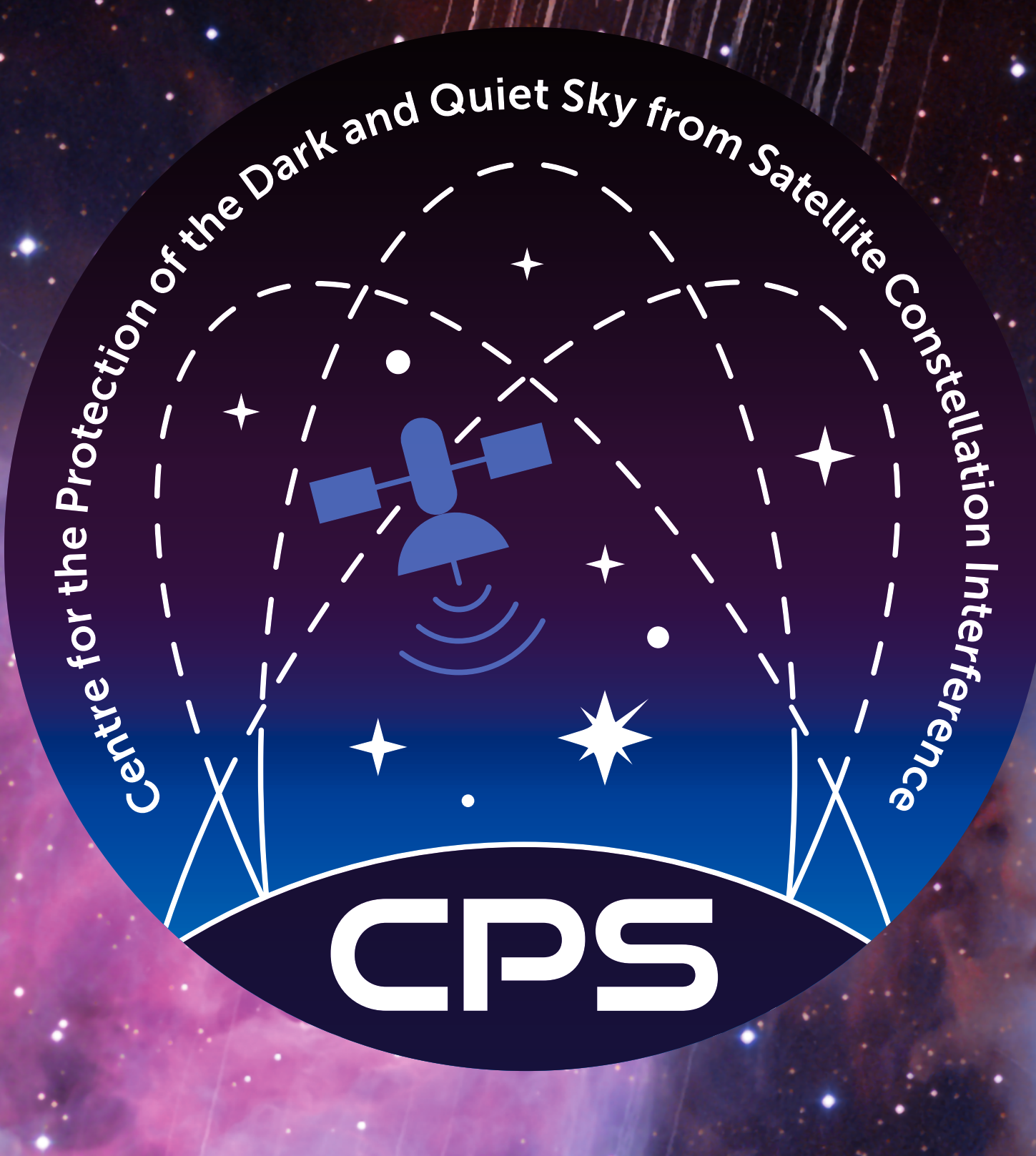

# Call to Protect the Dark and Quiet Sky from Harmful Interference by Satellite Constellations

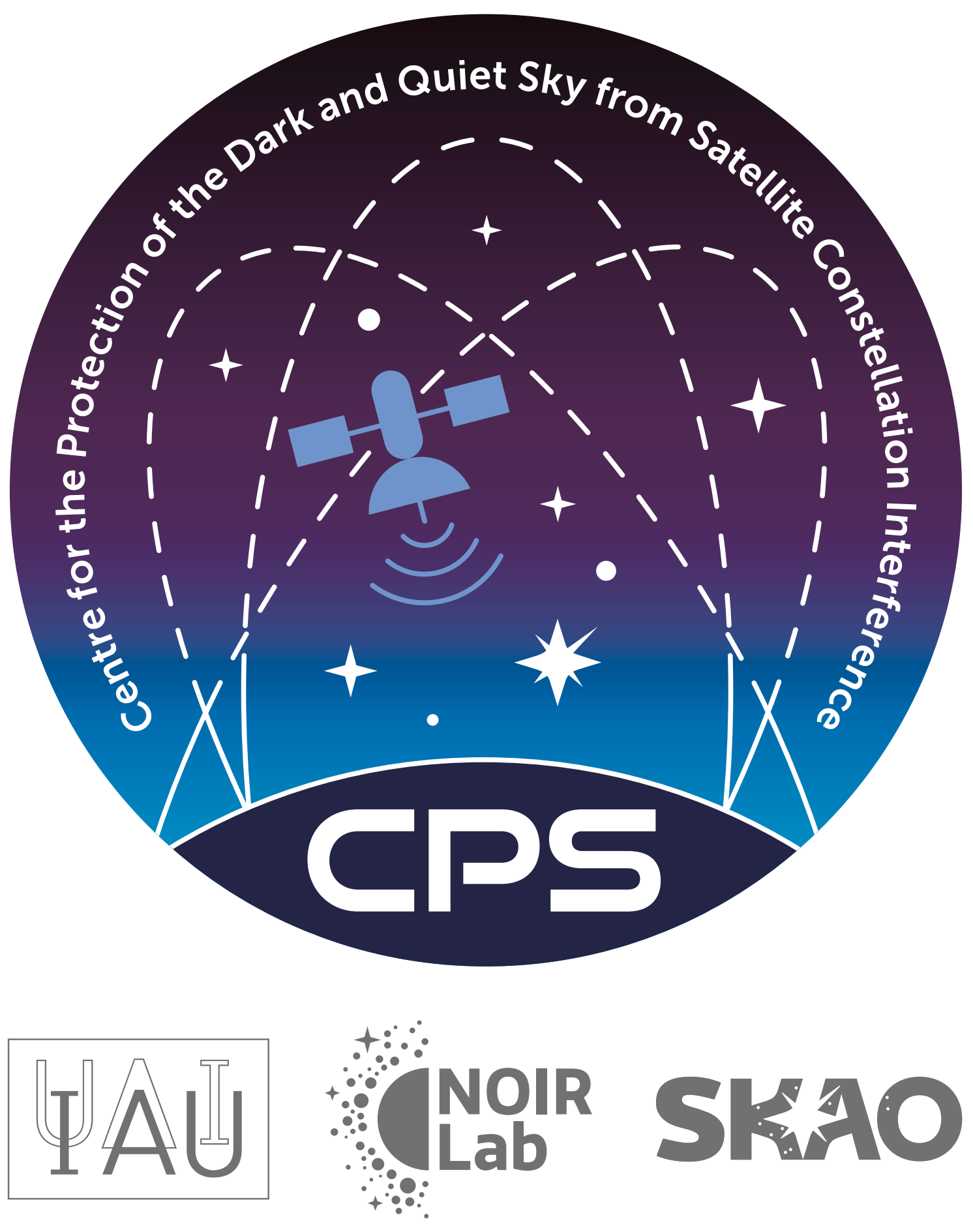

This document outlines the position of the International Astronomical Union (IAU) Centre for the Protection of the Dark and Quiet Sky from Satellite Constellation Interference (CPS) regarding the impact of low-earth-orbit (LEO) satellite constellations on scientific and public access to the night and radio sky. The intended audience includes government organisations, space industry, the public, and the astronomy community. While it does not delve into technical details extensively, it focuses on presenting the overall context, key concepts and recommendations.

The document comprises sections at three levels of detail, with the Executive Summary providing the briefest, the chapter Summary and Recommendations a more comprehensive, and the main body of the document a detailed explanation of the position of the CPS. As the situation with LEO satellites is constantly evolving, the CPS may issue an update to this position in the future.

To get in touch about this document,
please contact policy@cps.iau.org

The views expressed in this document are those of the IAU CPS and may not necessarily reflect the opinions of the individual contributors or the affiliated Institutes mentioned in the Acknowledgements.

# [ EXECUTIVE SUMMARY ]

The growing number of satellite constellations in low Earth orbit (LEO) enhances global communications and Earth observation, and support of space commerce is a high priority of many governments. At the same time, the proliferation of satellites in LEO has negative effects on astronomical observations and research, and the preservation of the dark and quiet sky. These satellite constellations reflect sunlight onto optical telescopes, and their radio emission impacts radio observatories, jeopardising our access to essential scientific discoveries through astronomy. The changing visual appearance of the sky also impacts our cultural heritage and environment. Both ground-based observatories and space-based telescopes in LEO are affected, and there are no places on Earth that can escape the effects of satellite constellations given their global nature. The minimally disturbed dark and radio-quiet sky[1] is crucial for conducting fundamental research in astronomy and important public services such as planetary defence, technology development, and high-precision geolocation.

Some aspects of satellite deployment and operation are regulated by States and intergovernmental organisations. While regulatory agencies in some States have started to require operators to coordinate with their national astronomy agencies over impacts, mitigation of the impact of space objects on astronomical activities is not sufficiently regulated.

To address this issue, the CPS urges States and the international community to:

1) Safeguard access to the dark and quiet sky and prevent catastrophic loss of high quality observations.

2) Increase financial support for astronomy to offset and compensate the impacts on observatory operations and implement mitigation measures at observatories and in software.

3) Encourage and support satellite operators and industry to collaborate with the astronomy community to develop, share and adopt best practices in interference mitigation, leading to widely adopted standards and guidelines.

4) Provide incentive measures for the space industry to develop the required technology to minimise negative impacts. Support the establishment of test labs for brightness and basic research into alternate less reflective materials and reduction of unwanted radiation in the radio regime for spacecraft manufacturing.

5) In the longer term, establish regulations and conditions of authorization and supervision based on practical experience as well as the general provisions of international law and main principles of environmental law to codify industry best practices that mitigate the negative impacts on astronomical observations. Satellites in LEO should be designed and operated in ways that minimise adverse effects on astronomy and the dark and quiet sky.

6) Continue to support finding solutions to space sustainability issues, including the problem of increasing space debris leading to a brighter sky. Minimising the production of space debris will also benefit the field of astronomy and all sky observers worldwide.

[1] In the following we refer to the radio-quiet sky as simply the 'quiet sky'.

# INTRODUCTION: LOW-EARTH-ORBIT SATELLITE CONSTELLATIONS AND ASTRONOMY

In recent years, numerous projects to deploy large 'constellations' of small satellites have been proposed by the satellite industry and by governments.

A satellite constellation is a set of spacecraft that share a common design, distributed across different orbits to provide a service and geographical coverage that cannot be achieved by a single satellite [1]. Applications of satellite constellations range from civil and military telecommunications to remote sensing and Earth observation, global navigation, and meteorology.

As of the end of 2023, there are more than 5000 satellites from large constellations in low Earth orbit (LEO) [2]. Filings in respect of over one million satellites at the International Telecommunication Union (ITU) indicate an ambition for possibly hundreds of thousands of satellites in LEO in the coming decade [3].

Astronomy today is conducted by almost 1000 largely publicly funded observatories and spacecraft, ranging from small metre-size telescopes to large-scale projects like the Square Kilometre Array and the Extremely Large Telescope, worth billions of dollars. Tens of thousands of astronomers around the world use such facilities to make scientific discoveries that benefit humanity. The professional community is complemented by an even larger amateur community, with commercially available equipment capable of supporting critical amateur–professional scientific collaboration. As a result of the ability to easily view the night sky and make important discoveries even with modest equipment, astronomy remains one of the most accessible sciences, particularly for developing countries.

The negative impact of artificial satellites on astronomy has been an issue for many years. However, the paradigm shift to an industrial use of LEO, with an expected 20- to 100-fold increase in the number of satellites, is becoming a much larger threat to science. The visibility and brightness of a satellite depend on the altitude of its orbit, its size, shape and surface reflectivity, its orientation with respect to an observer, and the orbital configuration of the system. LEO satellites reflect sunlight and can be detected as moving spots, particularly at twilight. Satellites at higher altitudes are visible for much of the night. They emit radio waves ubiquitously and can outshine cosmic sources by millions of times, which becomes an issue when they appear in direct line of sight to a radio telescope. The growth of LEO satellites has a direct impact on professional and amateur astronomy, and, without appropriate mitigations, it will have a significant and detrimental effect on the dark and quiet sky across the globe, even rendering some observations impossible.

Industry leaders have made substantial investments to reduce the optical and radio visibility of their satellites. They are entering into agreements with individual national governments to explore efforts to do so [4]. At the time of writing, the effort seems to be making progress. Measurements in 2023 of the brightness of SpaceX's Starlink Generation 2 mini-satellites have shown that the visibility in operational orbit with the specular coatings and darkening efforts is reduced compared to the satellites initially deployed in 2019, although bright Generation 2 satellites at final orbital height can still be observed [5]. There are many other satellite constellations in development, however, which will require brightness mitigations and partnerships with the astronomy community to validate those mitigations.

Satellite companies are developing various constellations to provide internet and telephony access, as well as Earth monitoring, some of which utilise dedicated ground-based antennas as user terminals. This growing novel and dense communication ground segment is in itself an added source of radio interference. More recently, there are growing efforts to develop systems that connect directly to common mobile/cell phone handsets. These satellites have a much larger surface area compared to those systems designed to connect to larger, less mobile ground stations. Without mitigation, these new satellites could outshine even the brightest stars and planets in the optical regime [6]. Currently, without voluntary cooperation, radio-quiet zones around sensitive radio telescopes are also not protected from satellite emissions, in particular direct downlinks. Moreover, the widespread transmission of radio signals, as well as unintended radiation in the radio regime in all directions, adds progressively more noise to the radio interference background [7].

# THE IMPORTANCE OF ASTRONOMY AND ASTROPHYSICAL RESEARCH

Astronomy plays an irreplaceable role in advancing fundamental physics and driving technological innovation in current and future industries. It serves as an accessible avenue for technical and scientific research and education in developing and developed countries. Earth-based astronomy is essential for providing the high-precision celestial reference frame needed for navigation and geodesy, monitoring potentially harmful space weather, and detecting and tracking potentially hazardous asteroids, comets and interstellar objects.

## SCIENTIFIC SIGNIFICANCE

Both historically and in the present, astronomical observations have driven breakthroughs in the understanding of the physical Universe and the mathematics required to describe it. Likely influenced by criticism of the Ptolemaic cosmology by the Persian and Syrian astronomers Al-Urdi, Al-Tusi and Ibn al-Shatir — all members of Maragheh observatory funded in what was then Mongolia, now Iran — Copernicus revived the idea that the Earth orbits around the Sun and not vice versa. Kepler's laws of the motions of the planets were the first mathematical description of a physical phenomenon, subsequently bolstered by Galileo's first observations of the heavens through a telescope. Newton developed a whole modern branch of mathematics to connect those motions to physical laws. All those giant leaps in scientific understanding depended entirely on direct observation of the cosmos under the dark sky. Einstein's general theory of relativity, which posits that mass changes the shape of space, was initially validated by observations of the apparent shift in the positions of stars close to the Sun during a solar eclipse. Today's observations of black holes and gravitational waves probe the fundamental nature of gravity itself. The existence of Dark Matter and Dark Energy, the main constituents of our Universe, has been inferred through astronomical experiments. Continued recognition of these critical contributions to fundamental understanding is exemplified by the fact that half of the Nobel Prizes in Physics awarded between 2017 and 2022 went to astronomers and astrophysicists.

Although we cannot predict the specific outcomes of today's fundamental research, history leads us to expect that it will significantly impact technological innovation and society in the coming decades and centuries. Furthermore, astronomy's aesthetic appeal often serves as a means to engage young individuals in scientific and technical subjects. The accessibility of astronomy, coupled with the allure of the night sky, offers opportunities for both developed and developing nations worldwide.

## TECHNOLOGICAL SIGNIFICANCE

Astronomy has driven technological innovations in imaging and image processing now found in many aspects of everyday life, including medical imaging, mobile phone cameras, telecommunication, airport scanners and high-sensitivity radio receivers.

The positive impact of astronomy on the development of new disruptive technologies and the economy is even more direct. For example, digital photography and WiFi technology have been pioneered in astronomy, as well as various medical image processing techniques. The next generation of observatories are driving innovations in 'big data', by creating techniques to process vast amounts of scientific data.

Fundamental discoveries are possible with small-aperture telescopes of modest cost. While LIGO made the first detection of gravitational waves, their origin was identified by a large number of small telescopes providing optical confirmation. Some of the first known exoplanets were discovered with telescopes smaller than 0.25 metres. Astronomy hence offers a low-cost entry point to fundamental physics and technological development.

## CULTURAL SIGNIFICANCE

Culturally, astronomy holds immense significance as it is, and has always been, deeply intertwined with the knowledge and daily lives of different societies across time. Celestial observation forms the basis of rich mythological and religious narratives across cultures. Archeological sites dating back to Neolithic and Chalcolithic times, to at least 6000 BCE, bear witness to connections to the sky, and numerous cultural artefacts depict stellar constellations. Ancient civilisations and many documented early human settlements developed astronomical observatories and built early astronomical knowledge for their daily lives, technology advances, and cultural and religious development. Major modern religions refer to the stars in their holy books, and many religious and cultural groups still depend on access to the night sky for wayfinding, ritual timekeeping, and more. Astronomical discovery was the origin of the fundamental transformation in western civilisation that led to the blossoming of exact sciences, which would later be called the Copernican revolution. Astronomical discovery permeates popular culture through compelling imagery as well as imaginative science fiction. Advertisements feature astronomical imagery, and commercial products are named after astronomical objects. Access to the natural dark sky is a key component of many cultural practices [8]; thus preserving and protecting a pristine night sky are crucial for safeguarding our cultural heritage.

## ECONOMIC AND SOCIETAL SIGNIFICANCE

Beyond fundamental research and technical development, astronomy has substantial economic implications.

Astronomy is an important component of many governments' space objectives and plans, along with other space-based activities, including the satellite communications and Earth sensing sectors. Ground-based astronomical observatories play a critical role in protecting our society's infrastructure and even life on Earth.

In order to provide navigation services for GPS receivers on Earth and also spacecraft operating in Earth orbit and beyond, we need precise knowledge of the Earth's shape and location. This is challenging, as Earth continuously changes its shape and orientation on various scales. We can measure minute shifts of Earth's shape and spin by knowing the positions of radio telescopes on the ground and the orientation of the Earth's rotation axis; this is done by observing a number of extremely distant radio sources (quasars), beacons of relativistic particles emanating from black holes. This forms the single most reliable reference system against which other navigation systems can be calibrated.

Observations of the radiation from these quasars with a terrestrial network of radio telescopes establish the basis for navigation, precision geolocation, continental drift measurements, climate change monitoring, and biodiversity assessment. These observations then require corrections supplied by the theory of general relativity — another discovery fundamentally supported by astronomy — for these precise applications. This service is currently a free contribution, financed by States, that provides an invisible foundational support to the worldwide economy.

Astronomical observations are also essential for monitoring space weather. Understanding the local space environment is vital for much of the modern economy, with infrastructure worth billions of dollars or euros at stake. Adverse space weather events can, for example, affect power grids on the ground and disable spacecraft in any orbit, potentially causing catastrophic cascades of destruction. With space weather monitoring, countermeasures can be initiated, and astronomy again provides a pivotal service to the worldwide economy.

Our planet has experienced multiple catastrophic impacts with asteroids and comets in the past [9]. Despite substantial international efforts, we still know only a fraction of the potentially hazardous objects that can enter Earth's atmosphere and cause local devastation [10]. Discovering unknown objects and monitoring the highly non-linear trajectories of near-Earth objects is the prime purpose of planetary defence, which relies primarily on optical telescopes.

Additionally, astro-tourism is developing a positive economic impact on rural communities in many countries [11]. Astro-tourists seek out sites with minimum impact from ground-based light pollution, desiring unhindered views of the natural night sky.

# THE IMPACTS OF LEO SATELLITE CONSTELLATIONS ON ASTRONOMY

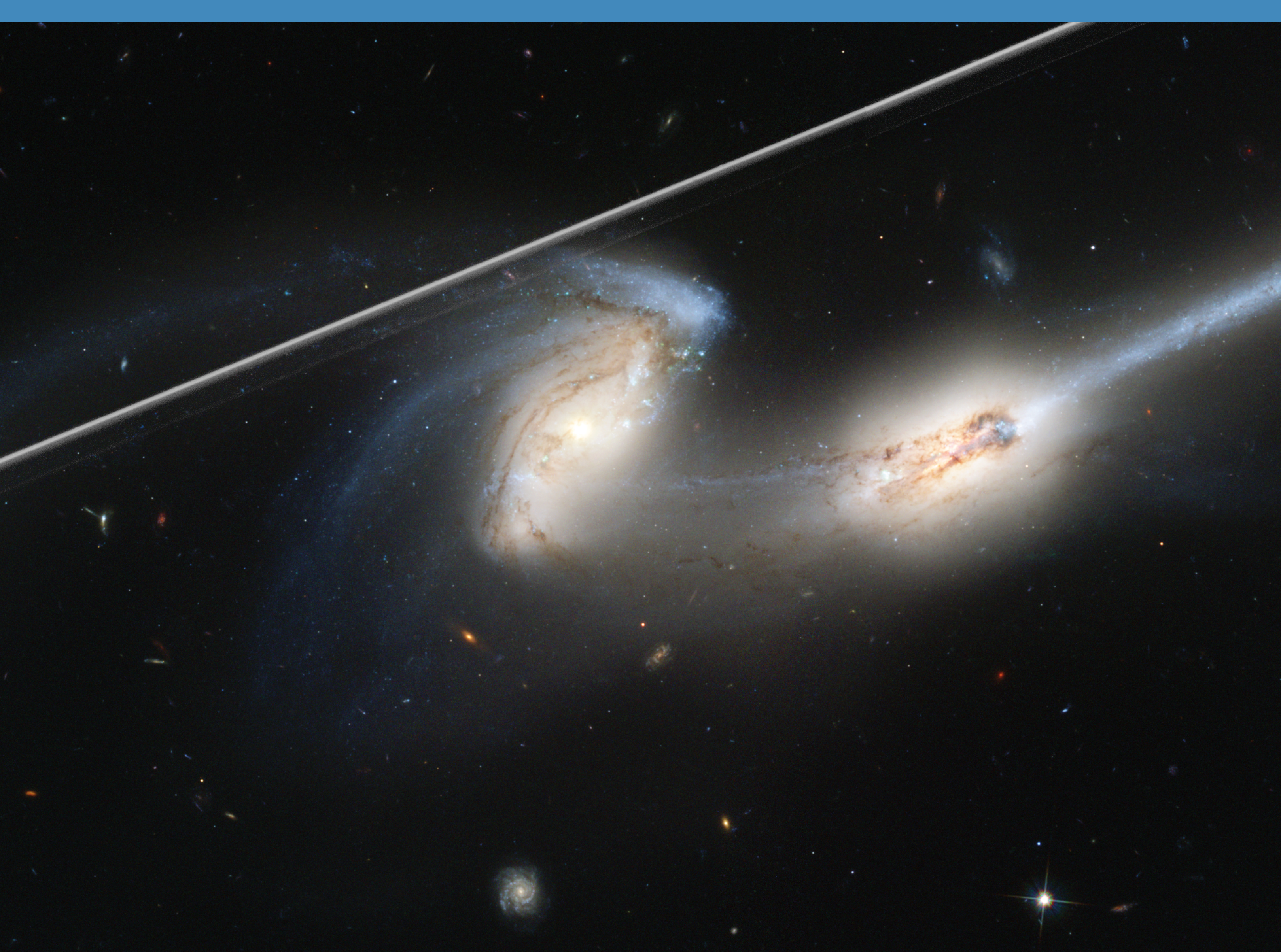

Figure 1: Bright streak left in an astronomical image of a group of galaxies with the Hubble Space Telescope. The streak was generated by a satellite crossing the field of view during the exposure. *Courtesy of NASA/STScI [13]*.

Satellite-induced light and radio pollution affects the night sky globally. This environmental impact means that a pristine night sky is being affected everywhere in the world by satellite reflections and space debris. Even traditionally remote observatory locations will be affected just as severely as any other area.

## VISIBILITY OF SATELLITES IN THE SKY

Most satellites in LEO are constructed of materials that reflect sunlight and as a result they can often be seen by the unaided eye. Depending on the satellite's altitude, orientation, and surface characteristics, and the darkness of the viewing site, these reflections can be observed throughout the entire night. The larger satellites that are intended to communicate directly with cellular phones are expected to be among the brightest objects in the night sky if mitigations are not implemented.

Efforts have been made by industry to identify and develop novel approaches to spacecraft design and mechanisms to reduce the brightness of satellite systems. However, the technology to reduce brightness is still in its early stages and is not yet broadly adopted by all constellation operators and consequently many systems already deployed are often visible to the naked eye in a moderately dark sky. The proposed and deployed LEO satellite systems are causing changes in the appearance of the night sky worldwide, impacting the cultural heritage associated with it in every country and every location on Earth [12].

## IMPACT OF REFLECTED SUNLIGHT ON OBSERVATIONS

LEO communication satellites, at their current levels of brightness, will adversely affect optical and infrared astronomical observations by reflecting sunlight.

As satellites move across the sky, they can reflect sunlight and leave bright streaks on astronomical images (see Figure 1). Depending on the brightness, such interference has the potential to invalidate whole datasets. Satellites in lower orbits enter the Earth's shadow earlier after sunset, ceasing to reflect sunlight to the ground and disappearing from view, at least from the perspective of an optical observer. Generally, the lower the orbit, the shorter the period of visibility after twilight, although the visible period also depends strongly on the geographical latitude of the observatory and the season. Orientation, size and the reflectivity of the materials it is made of also play important roles in determining the brightness of a satellite as seen by astronomical observers.

Depending on the field of view of the telescope (the portion of the sky it observes), it might not be possible to avoid the negative impact of streaks of reflected sunlight on professional and amateur astronomical observations, particularly in twilight. Studies show that the extent of the impact depends on the telescope's characteristics, science plans, and observing strategies. For instance, simulations that assume some 60 000 LEO satellites in realistic orbital configurations predict that Vera C. Rubin Observatory, hosting a large optical telescope with a wide field of view that is scheduled to commence operations in 2024, may be impacted by at least one satellite streak on up to 30% of observations at the beginning and end of each night [14]. Current asteroid population models suggest that unknown potentially hazardous objects reside in greater numbers in locations that are accessible only through observations conducted during twilight, shortly after sunset or before sunrise. But it is exactly during twilight that the sky is most affected by LEO satellite constellations [15].

The combined effect of current satellites in orbit and the accumulation of space debris may already be causing an approximately 10% increase in the brightness of the night sky compared to natural levels [16]. This exceeds the classification threshold established by the International Astronomical Union (IAU) in 1979, which should not be surpassed at large astronomical observatory sites. Consequently, the effective sensitivity of optical and infrared astronomical telescopes on Earth may have already decreased as a result of the unprecedentedly high activity in LEO.

Any increase in space debris resulting from the growing use of low Earth orbits also impacts astronomy. Sunlight reflecting off debris brightens the night sky in a similar way to nearby city lights increasing the level of light pollution. Even small reflections from uncontrolled objects in LEO will also cause false detections of transient events of astrophysical interest [17].

In addition to constellations of many smaller satellites, there is a growing interest in larger structures in space, including space-based solar power systems, LEO space habitats, and larger satellites that connect directly to mobile phone handsets. Without countermeasures, all these plans would result in sources of even greater interference to astronomy.

Unmitigated LEO satellite systems, with a steadily increasing density in the sky, are hence a threat to optical and infrared astronomy, and therefore to a fundamental driver of science and technology

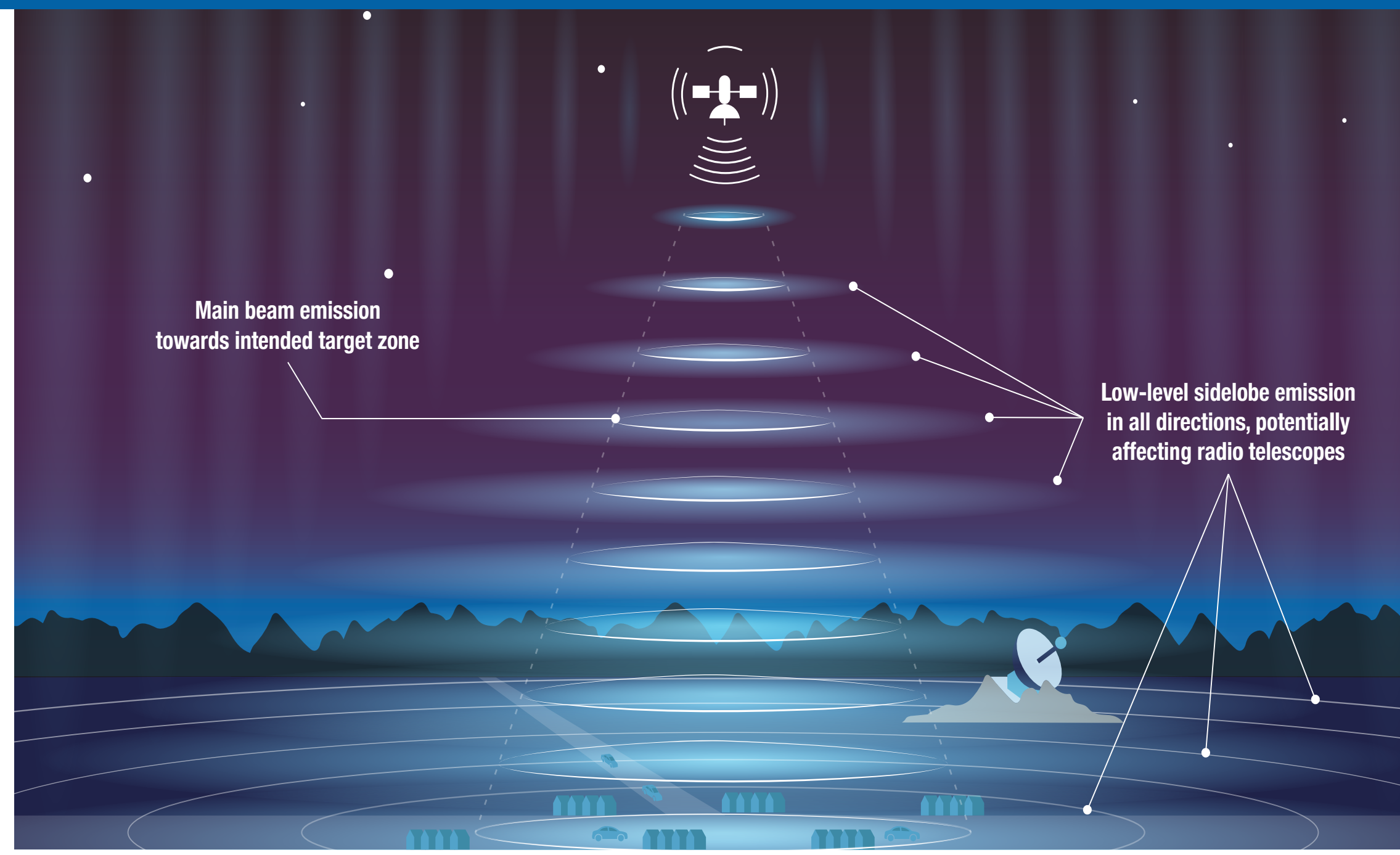

Figure 2: Illustration of the illumination pattern of antennas of satellites in large constellations. The radiation is emitted not only in the intended direction, but also at lower power level in a much wider cone, at some level even in any direction. Since radio telescopes are, to some extent, also susceptible to radio emission from any direction, they detect emission from all satellites that are above the horizon. *Image credit: IAU CPS/NOIRLab/SKAO.*

## RADIO ASTRONOMY

As in the optical, satellites can also reflect radio emissions from the Sun or terrestrial anthropogenic radio signals towards radio telescopes, resulting in a reduced quality of observations and the loss of data. Far more significant, however, are the active radio transmissions by the satellites and the unnecessary leakage of radiation from on-board electrical equipment and electronics.

Even a perfect radio transmitter will always produce some emissions at frequencies outside the designated carrier signal's band. These can be reduced with filters or other techniques, but only to a certain level. Better suppression is usually tied to increased effort and thus costs. The same is true for the emissions of an antenna. While most of the power is beamed towards a certain area — the main beam — there is always a fraction of the power that goes into other directions (see Figure 2). This means that any transmitting satellite visible to a radio observatory will cause a signal with the potential to disrupt radio observations, even at frequencies other than the nominal carrier frequencies of the satellite. The strength of the detected signal depends on the pointing orientation of both systems. As they move over the sky, satellites can appear in the direct line of sight of a radio telescope, which will amplify their emissions by a huge factor compared to terrestrial transmitters, which are usually located at (or behind) the horizon. Furthermore, even in the most remote observing sites, far from terrestrial transmitters, satellites will still be high in the sky and will be a potential source of harmful interference. This results in distortions of astronomical measurements, from sub-noise-level features that can become apparent with long observing times (e.g., multi-year observations of the cosmic microwave background) to the loss or prevention of entire datasets. In extreme cases, spaceborne Earth or weather exploration radars can damage radio astronomical receiver systems [18]. Satellite emissions can also mimic or interfere with observations of transient astronomical radio sources.

Large satellite constellations hence pose a challenge to the traditional operation of radio observatories. Many observations make passive use of a wide range of frequencies outside of those explicitly allocated for radio astronomy by international regulation. This includes bands explicitly allocated for use by satellite systems. Radio astronomy's passive observation in these bands does not interfere with other uses of the spectrum. However, observations at these frequencies are affected by active transmissions by other users and there are no official rights of protection from interference from authorised services.

These opportunistic observations of cosmic radio sources in parts of the spectrum not allocated to radio astronomy are in many cases made possible thanks to national or regional protection zones, as well as geographical separation and natural shielding against terrestrial anthropogenic radio emissions. However, no existing or potential radio observatory sites benefit from such protection zones in the case of satellite constellations, which are designed to transmit signals to everywhere on Earth. No international regulation exists to protect those zones or radio observatories outside the limited bands allocated to radio astronomy.

Furthermore, the aggregate unintended radiation or leakage radiation emitted by electronic systems on satellites is less regulated and can occur in bands allocated to radio astronomy. This leads to, at best, a significant loss in sensitivity and may severely affect certain ground-based radio observations [7].

# PROPOSED MEASURES AND MITIGATION STRATEGIES

## ASTRONOMY COMMUNITY MITIGATIONS

Most data for astronomical and planetary science research come from ground-based observatories, as do those used for planetary defence and a census of orbiting objects. In addition, major observatories are funded by international consortia; there are currently 40 large ground-based optical telescopes with primary mirrors between 3 and 11 metres in diameter, representing a worldwide investment. In contrast, there is only one telescope this size in space. Facilities in space cannot replace ground-based telescopes, owing to a wide variety of prohibitive technical and cost challenges [19].

Astronomers are working to mitigate data loss due to satellite constellation interference in the optical/infrared and in the radio regime.

Support is required for the development, deployment and maintenance of software packages to identify and mask out satellite streaks. Accurate predictions of the positions of satellites on the sky in a given direction and at a given time of night are essential. Those can accelerate identification of streaks in wide-field images without requiring a blind search of the full image. Some satellites will be so bright that they must be avoided on the basis of such predictions, in order to prevent the complete loss of an image. Development is required for applications to make access to operator-supplied position data routinely available to observatories. Coordinated observations of the apparent brightness of constellation satellites provide feedback to operators on the effectiveness of their efforts to dim the spacecraft and can feed into the predictions to provide both position and brightness as they would impact a planned observation. The IAU Centre for the Protection of the Dark and Quiet Sky from Satellite Constellation Interference (CPS) provides the organising structure, but continuing support to observatories around the world is required.

Unwanted effects by human-made radio transmissions have affected radio astronomy since its inception. Accordingly, the astronomical community has always had to put significant resources into the implementation of various countermeasures, a long-standing, ongoing effort. Three methods can be used. The simplest is to avoid the impeding signal. However, the possibility of minimising direct observations of satellites at radio frequencies by sophisticated observation strategies is more limited than in the optical regime, as radio antennas are susceptible to radio emission from the entire sky. Even if the sensitivity decreases rapidly beyond the pointing direction, it never vanishes. A second method, subtracting the interfering signal from astronomical observations, has been attempted, but with very limited success. To achieve good results, each interfering signal should be observed to a better accuracy than the astronomical source, which is virtually impossible. The third method is the removal of both the astronomical signal and the interfering signal when the latter occurs, and then restricting any analysis to the remainder of the data. The task is to identify the false signals from interfering sources, and several techniques have been developed. They can be slightly optimised if the satellite's position is known well enough, again an effort being undertaken in the scope of the CPS.

All mitigation methods have in common that the degradation of the astronomical signal cannot be avoided, ranging from an increase of the noise against which a signal needs to be detected, over false detection of astronomical sources, to the total loss of the observation.

In both the optical and the radio regime, further resources are needed by astronomers and the satellite community to develop new, and optimise existing, mitigation strategies and techniques so that they are affordable, accessible and effective. Many governments actively support the development of space activity broadly, including the use of satellite-delivered services, satellite launch and manufacturing and investments in ongoing scientific discovery in optical and radio astronomy. Considering the interdependencies of the space ecosystem and the consequences for astronomy and space sustainability generally, mitigation then becomes a whole-of-government balancing challenge, and ongoing support from government is necessary to find and implement effective solutions for all parties concerned to coexist.

## COLLABORATION WITH INDUSTRY

Satellite constellation systems are being launched and deployed more rapidly than before, as a result of more efficient spacecraft manufacturing, improved access to launch facilities, competitive market forces, and international and national regulations that require a certain percentage of a constellation to be deployed within specified time periods. As a result of this rapid deployment and the absence of binding regulations to mitigate impacts on astronomy, the IAU and the most impacted observatories have prioritised direct collaboration with operators to explore and develop mitigations in this new field of technology. This collaboration has already produced several innovative solutions and ideas, which some operators are already implementing and sharing with the community.

Industry leaders have pledged to make, and have made, substantial investments [20] to reduce the optical and radio visibility of their satellites. Several constellation operators have begun applying these best-practice mitigations voluntarily, and certain governments are now requiring their licensees to enter into agreements to explore mitigations in collaboration with the astronomy community. Despite this work, at the time of writing there is no mitigation approach that has successfully reduced the brightness of a LEO satellite to the quantitative target level recommended by the astronomy community (see Box 1). Further work is required to explore other mitigation options available to satellite constellation operators and predictive tools. Government support is needed to establish a network of test labs that can predict brightness from prototype satellites, as is available for radio frequency testing, and also basic research to assess alternative materials that may reduce brightness.

The CPS encourages industry to continue collaborating with the astronomy community and encourages governments to find ways to facilitate and support this collaboration: developing brightness mitigations; sharing precise positional data to coordinate with observations; implementing suppression measures of unwanted radiation; and other measures. The CPS also encourages companies to work together and share solutions to support astronomy, and encourages States to put in place incentive measures to help satellite operators to develop the required technology to achieve maximal mitigation of interference with astronomy and the dark and quiet sky. Companies and States are also encouraged to collaborate with their national astronomical societies and the CPS, which has established an Industry and Technology Hub specifically to encourage industry collaboration and sharing of mitigations.

## VOLUNTARY CODES OF CONDUCT

The space sector is actively reviewing, updating and initiating best practices, voluntary codes of conduct and regulatory norms to support a space ecosystem that features multiple large satellite constellations, as well as other novel space activities. Examples include the Inter-Agency Space Debris Coordination Committee (IADC) space debris mitigation guidelines, the Space Safety Coalition's Best Practices for the Sustainability of Space Operations, the Long Term Sustainability Guidelines developed by the United Nations Committee on the Peaceful Uses of Outer Space (UN COPUOS), the Code of Conduct for Space Sustainability of the Global Satellite Operators Association (GSOA), and the ESA Zero Debris Charter. These guidelines related to space sustainability offer space actors the flexibility to adopt best practices and promote responsible behaviour without imposing rigid constraints, and serve to inform governments in the development of policy and regulation in the longer term.

The CPS encourages space sector actors to incorporate mitigations for reducing impacts on the dark and quiet sky and astronomy as part of their consideration of good corporate citizenship and environmental social governance (ESG) principles, and to participate in public systems such as the Space Sustainability Rating (SSR) [21].

The SSR, as one example of these public systems, provides a new and innovative way of addressing orbital challenges by encouraging responsible behaviour through a rating system informed by a transparent, data-based assessment of the level of sustainability of space missions and operations. The SSR system supports space actors, such as governments, space agencies, and commercial companies, in understanding the impact of their activities on the space environment, and identifying opportunities to minimise those impacts.

In 2022 the development of a Dark and Quiet Sky module was initiated in the SSR to enhance the rating system, with the support of the CPS Policy Hub [22]. The primary objective of this module is to develop a quantification methodology to evaluate the impact of satellites on astronomical observations, for both optical and radio astronomy.

## REGULATORY MEASURES

Analysis of international space law documents, in particular the Outer Space Treaty (OST) [23], and the relevant discussions in international fora (e.g., UN COPUOS) demonstrate that astronomy constitutes a lawful form of accessing, exploring and using outer space. Likewise, the deployment of satellite constellations is a lawful form of using outer space, and operators have a legal right to transmit signals in their assigned frequency bands as determined by the spectrum management processes convened under the ITU. Unlike the ITU system, which allocates different portions of the electromagnetic spectrum to different users, including radio astronomy, the OST confers no hierarchy of priority for space activities; accordingly, balance and coordination must be found between astronomical observations and satellite constellation activities.

Many States are developing new regulatory approaches to approve and supervise national space activities, including satellite constellations, as part of their obligations as parties to the OST and as members of the ITU, and in compliance with general principles of international law. Regulations typically focus on launch safety, spectrum allocations and usage, and mitigations of space debris and conduct of safe on-orbit operations. With the exception of the limited, pre-existing protections for radio astronomy under the ITU and in some instances by implementing protection zones under national law, and a more recent practice by the US Federal Communications Commission (FCC) to require non-geostationary satellite operators to reach coordination agreements relating to science impacts with NASA and the US National Science Foundation, States do not presently consider the impacts on ground- or space-based astronomical sites.

There are four domains in which the CPS encourages States to explore the development of regulatory mitigations: spectrum management, environmental law, cultural protection and regulatory supervision.

## SPECTRUM MANAGEMENT

The ITU Radio Regulations (RR) [24] are maintained and regularly updated by the ITU, the oldest UN agency. The RR have the status of an international treaty and aim to regulate the use of the radio spectrum internationally to prevent harmful interference between different radio services, considering that radio waves can propagate over long distances. Radio astronomy is recognised as one service among others. This recognition reflects the significant investments made by countries in constructing and operating radio observatories and the associated infrastructure. Consequently, the RR define different levels of protection for radio astronomy in specific frequency bands, ranging from binding international protection to recommendations to protect radio astronomy in national law. National regulators adopt the RR into national law and police them, including the mandatory protection of astronomy. As the spectrum management process has been well established over many years, a number of spectrum management organisations and radio observatories advocate for radio astronomy and interface with national and ITU spectrum bodies.[2]

However, while offering some shelter, the existing regulations require a substantial update in several aspects to sufficiently protect radio astronomy in the era of large satellite constellations.

The first aspect is a regulatory measure, which can be addressed within the ITU. Adherence to the regulations protecting astronomy in its allocated bands is neither checked centrally by the ITU nor structured to take aggregate effects into account through international forward planning. While a simple limit to the interference from single constellations exists (2% loss of data), and a simple limit to the interference from all satellite constellations exists (5% loss of data), no procedure is in place that regulates how the burden to limit the aggregate impact is shared among constellation operators. This may lead to a situation where the aggregate signal of multiple satellite constellations from different States transgresses protection levels for astronomy in an uncontrolled way.

Secondly, the RR protect only a small portion of the radio spectrum used for radio astronomical measurements. Some States therefore implement additional radio-quiet zones around radio observatories, in which radio emissions have to be suppressed within selected bands or across the whole radio spectrum, not necessarily protected by international regulation. However, the current radio-quiet zones defined at the national level do not account for satellite emissions, and radio telescopes lack protected access to frequencies beyond the small portion allocated to radio astronomy.

Thirdly, while there exist international electromagnetic compatibility (EMC) compliance standards to regulate leakage radiation from devices operated on Earth's surface, no international equivalent has yet been introduced for spacecraft. [7].

The effect of the reflection of radio signals by satellites from the Sun and ground-based sources requires further study.

The CPS recommends that States:

- Observe and strengthen the protection of radio astronomy from satellite constellations in radio bands allocated to radio astronomy by developing appropriate measures in the ITU Radio Regulations.
- Recognise that modern radio astronomy requires broadband spectrum protection beyond the allocated narrow frequency bands in the ITU Radio Regulations, and hence that national and international regulatory protection from large satellite constellations is also required to preserve and create specific geographic areas designated as radio-quiet zones around telescopes.
- Develop and adopt a global electromagnetic compatibility (EMC) standard for unintended radiation from the electronics of satellites in large constellations which recognises the protection of radio astronomy.
- Support studies to investigate the impact of reflections of solar radiation and of terrestrial signals from satellites in large constellations that can impact on radio astronomy.

## ENVIRONMENTAL LAW

Article III of the OST specifies that activities in the exploration and use of outer space should be carried out in accordance with international law, part of which includes environmental law. A number of legal instruments exist that could be applied to the issue of unintended pollution from satellite constellations in the context of light pollution and unwanted radio emissions. For example, the 1979 Geneva Convention on Long-Range Transboundary Air Pollution uses a broad definition of pollution, which includes energy released via human activities. Dark and quiet sky impacts add to a growing number of issues associated with uncontrolled satellite constellation activities, including broader ecological and environmental impacts [30].

[2] Scientific Committee on Frequency Allocations for Radio Astronomy and Space Science (IUCAF) [25]; Committee on Radio Astronomy Frequencies (CRAF) [26]; Square Kilometre Array Observatory (SKAO) [27]; International Astronomical Union (IAU); Committee on Radio Frequencies (CORF) [28]; Radio Astronomy Frequency Committee in the Asia-Pacific region (RAFCAP) [29].

The CPS encourages States to:

- Consider how tenets of international environmental law, such as the precautionary principle, the 'polluter pays' principle, the prevention of transboundary harm, and the principle of sustainable use should be applied to activities in outer space, and particularly to those taking place in LEO [31].

### CULTURAL PROTECTION

Some countries make provision to protect the night sky from the impact of light pollution from terrestrial sources. There is a rich diversity among peoples throughout the world who have societal, economic, and cultural interests in the exploration and use of outer space, including dark and quiet skies.

The CPS encourages States to:

- Find mechanisms to recognise the night sky as part of our common cultural heritage and include its protection from satellite interference.

### REGULATORY SUPERVISION

When States licence space launches, whether public or private sector, they have a responsibility under the OST to authorise and continually supervise the activities of spacecraft for their complete duration, from launch, over operation, to de-orbiting. States must also exercise due regard to the on-orbit operational phases of satellite operations and the potential impact they may have on the environment and society. States are also responsible for regulating private actors to ensure their compliance with international law and principles through the imposition of relevant requirements for the authorisation and supervision of space activities.

For the visible spectrum, international regulation has not been seen as a requirement, since ground-based lighting seems to impact only the regional environment, although its impacts can extend beyond national borders [7]. Historically, light pollution has been addressed through national or more local regulation. Optical communication systems and ground-to-space lasers are less prone to interference, thus making coexistence questions less important than in the radio domain. However, large satellite constellations and other emerging projects in LEO, with optical reflections affecting astronomical infrastructure across national borders, require a paradigm change.

The CPS recommends that government authorities responsible for the authorisation and supervision of the launch and operation of satellites require an impact assessment, with as much public information as possible, and base their subsequent licensing decision on criteria covering but not limited to:

- An analysis of brightness with the best tools available, and the guidance of the industry, taking into account the quantitative limits defined in this document;
- An analysis of the cumulative radio emissions from intended emission and unintentional electronic noise, including strategies to minimise the impact on radio-quiet zones;
- Requirements for operational positional data sharing to mitigate impacts on astronomy;
- Consideration of the balance between mission objectives, orbital sustainability, and interference on other space activities;
- Requirements for careful, interdisciplinary environmental assessment.

Furthermore, governments should continue to find ways to bring the matter of the dark and quiet sky to the attention of the UN COPUOS and its sub-committees, and work towards bilateral or multilateral agreements as a stepping stone towards international regulation.

## TECHNICAL MEASURES

Several measures to ensure space sustainability are also relevant for astronomy. Policy makers and industry are encouraged to implement these.

To support the dynamic growth in LEO projects, governments are escalating their Space Domain Awareness (SDA) activities to identify, track and monitor spacecraft and debris to aid in collision avoidance and increase space situational awareness. The introduction and further development of an international, publicly available, electronic orbital register for LEO satellites, with rapidly updated orbital parameters, will allow astronomical observatories to predict the occurrence and locations of satellites in the sky and may be used to mitigate their negative impact by avoiding observations of fields with a high density of satellites. Interest in collision avoidance and debris mitigation is already driving the development of private-sector and governmental systems for SDA and traffic management. Space debris can affect observations similarly to satellites themselves. In some instances, the fidelity of data required by astronomers may exceed that required by satellite operators for their on-orbit collision avoidance.

The CPS encourages States to:

- Consider astronomy requirements when developing SDA programmes and to increase the accuracy of space debris location and channel that information into astronomy databases.
- Obligate satellite operators to assess the impact of their system on astronomy, to share space location information with internationally-available SDA platforms, and to adopt the prevailing best practices and available technical measures to protect astronomy.
- Support the evolving tools and technology for brightness mitigation by establishing test labs to predict brightness from prototypes and to perform basic research in less-reflective alternative materials for spacecraft fabrication.
- Support the development of techniques to suppress radio emission in unwanted directions and at unwanted frequencies from satellite transmission systems.
- Support the development of techniques to suppress unwanted electromagnetic radiation from satellites.

As space debris will affect astronomy by increasing background brightness [16], measures to avoid its increase will also benefit astronomy. These measures should be geared towards avoiding an increase in background brightness by more than 10% above the natural levels. The short-term allocation of orbits and the enforcement of fast and reliable deorbiting strategies are therefore of great interest to astronomy. In addition, as constellation operators improve their technological approaches to lowering brightness and limiting unintended radio detectability, the expectation is that they will incorporate these improvements into the satellites if replacement is needed.

In addition to the general space-sustainability rules, the CPS encourages administrations to implement regulations to specifically protect astronomy and the night sky. The most urgent measures are listed below.

While the presence of satellites during astronomical observations will always have a detrimental effect, the worst (loss of whole exposures or damage to the electronics) can be prevented if satellites in the lowest orbits are not visible to the naked eye and if they become dimmer with increasing altitude. The latter requirement is a consequence of a reduced angular speed at increased altitude; the single affected receiver pixels are exposed for longer to the satellite reflections, and therefore a dimmer satellite has the same detrimental effect. The CPS has formulated comprehensive technical criteria for adoption. The most important thing for optical astronomy is to limit the brightness of satellites, as given in Box 1.

Another important concern for astronomy is the visibility of satellites during the darkest part of the night. For the specific criterion of limiting that visibility for the locations of the world's largest research telescopes, the recommendation is to have constellation orbits that stay below roughly 600 km [32].

To better protect radio astronomy, high-quality transmission systems should be used to illuminate only targeted cells on

**Box 1: The IAU recommendation to limit satellite brightness**

The IAU recommends a visual brightness limit for LEO satellites in operational orbit: satellites should never be visible to the naked eye, and the maximum allowed brightness should be lower with increased altitude. Although satellites do appear fainter when they are farther away from the Earth, that effect is somewhat offset by their apparently slower movement across the sky at greater altitudes. This is expressed in the formulas

$$V_{mag} > 7.0 \quad \text{if SatAltitude} \leq 550\,\text{km},$$
$$V_{mag} > 7.0 + 2.5 \times \log_{10}(\text{SatAltitude} / 550\,\text{km}) \quad \text{if SatAltitude} > 550\,\text{km},$$

for both photopic magnitude as well as Johnson *V* magnitude m. The magnitude $V_{mag}$ is a measure of the brightness in the visible regime, where photopic is visually perceived and the Johnson system is filter-defined and calibrated. SatAltitude is the altitude of the satellite above sea level. A higher value of $V_{mag}$ corresponds to a dimmer object. Figure 3 illustrates the acceptable brightness limit.

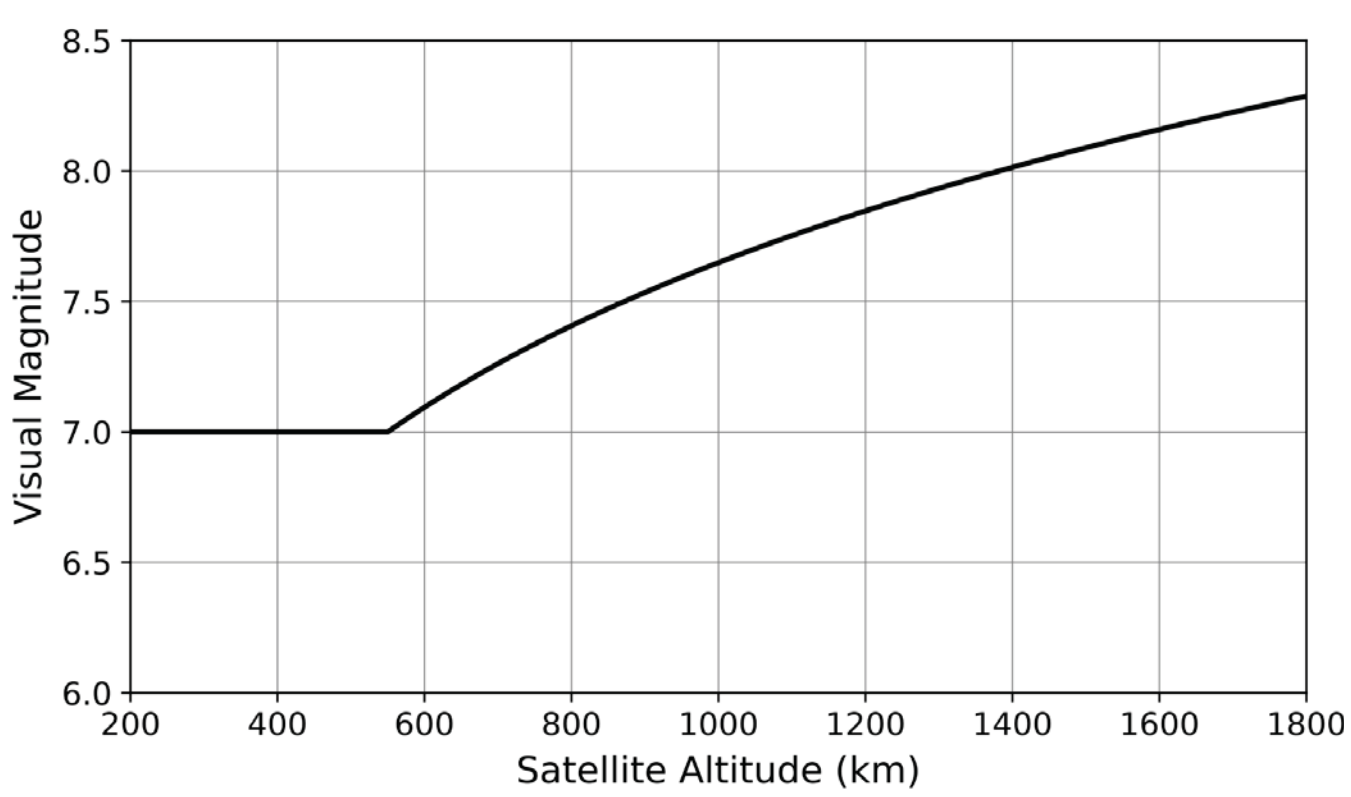

*Figure 3: Brightness limit of LEO satellites as proposed by the IAU. The vertical axis denotes $V_{mag}$. [31]*

the ground without spilling emission into their surroundings, as well as effective filters to restrict the radio emission to the bands used for the radiocommunication, in order to avoid unwanted emission in frequency bands used for radio astronomy. Leakage radiation from non-transmitting parts of the satellites should be suppressed effectively.

The CPS encourages States to oblige satellite operators to enact technical measures to protect astronomy. In particular, reflections of sunlight towards optical telescopes should be avoided, and technical measures should be implemented to constrain emission spatially and in frequency. Rules for electromagnetic compatibility of spacecraft with terrestrial equipment (radio telescopes) are required, such as that the aggregate unwanted radiation from satellites stays at a level that does not entail harmful interference to radio astronomy. Protection zones for astronomical facilities extending into space should be considered.

# [ SUMMARY AND RECOMMENDATIONS ]

## SUMMARY

In recent years, the cost of access to space has been reduced significantly, opening the doors for full commercial utilisation. Commercial satellite companies and governments have endeavoured to launch large-scale space missions with the aim of supplying individual, commercial and government customers worldwide with broadband internet, Earth-observation data, and other applications. While undoubtedly improving life for many customers, there are some negative consequences. Astronomy and the uninterrupted access to the night sky are among the most affected fields.

This document is addressed to government administrations, the space industry, the public, and astronomers and presents a summary proposal to implement mitigations to reduce and eliminate the impact of low-Earth-orbit (LEO) satellite constellations on scientific and cultural access to the night sky. It is not meant to cover the detailed discussion of technical aspects, but highlights the context and the most important concepts.

- LEO satellites, deployed in large numbers, reflecting light from the Sun as well as being visible to infrared and radio detectors, have become a threat to the dark and quiet sky and will have negative impacts on astronomy, fundamental science, and wider society.

- The dark and quiet sky is a requirement for fundamental research and technology development, and programmatic observations contributing to key priorities like planetary defence, and is essential for basic services like geolocation.

- Without countermeasures:

    - LEO Satellites will appear as numerous moving objects in the night sky, visible to the unaided eye much of the time, effectively transforming the view of the stars;

    - LEO Satellites will have a significant and, in some cases, severe impact on astronomical observations, including those searching for potentially hazardous asteroids or comets and those used to calibrate geolocation and observations used to calibrate geolocation;

    - Radio transmissions and unintended electromagnetic radiation from satellites will have a significant and in some cases severe impact on radio astronomy.

- Large LEO constellations hence compromise the effectiveness of astronomical infrastructure and are detrimental to national investments in astronomy, scientific discovery, technical spinoff, and fundamental services to society, scientific discovery, technical spinoff, and fundamental services to society..

- The undisturbed dark and quiet sky is part of our cultural heritage that requires and deserves protection.

## RECOMMENDATIONS

- The CPS encourages States and the international community to make every practical effort to prevent the catastrophic loss of high quality observations of the dark and quiet sky through establishing regulations and terms of licensing that require operators to exercise due regard for global visibility and to mitigate the impacts on observational astronomy through:
  - A visual magnitude limit as described in Box 1, ensuring the invisibility of LEO satellites to the naked eye and mitigating some data loss in observations, initially on the basis of best efforts until technology is mature;
  - Addressing optical visibility over the full mission lifetime, including orbit raise and de-orbit;
  - Implementing and policing existing international regulation and strengthening the protection of radio astronomy from satellite constellations in radio bands allocated to radio astronomy by developing appropriate measures in the ITU Radio Regulations and on a national basis;
  - Recognising that modern radio astronomy requires broadband spectrum protection beyond the allocated narrow frequency bands in the ITU Radio Regulations, and hence that national or international regulatory protection also from satellite constellations is required to maintain the efficiency of radio quiet zones around telescopes;
  - Introducing a global electromagnetic compatibility (EMC) standard for unintended radiation from the electronics of satellites, which recognises the protection of radio astronomy;
  - Proper implementation of the relevant tenets of the OST and its elaborating conventions and the ITU Radio Regulations;
  - Consideration of the applicability of the environmental legal framework to outer space and activities conducted therein, for the purpose of protecting astronomy and mitigating impacts on Earth's ecological systems;
  - Space sustainability laws limiting the generation of space debris and facilitating a fast deorbiting strategy, with the purpose of limiting the global increase in diffuse sky background to no more than 10% artificial contribution;
- The CPS recommends that government authorities responsible for licensing the launch and operation of satellites require an impact assessment and base their subsequent licensing decision on criteria covering but not limited to:
  - An analysis of brightness taking into account the quantitative limits defined in this document, based on well-measured materials reflection properties;
  - An analysis of the cumulative radio emissions and unintentional electronic noise, including strategies to minimise the impact on radio-quiet zones and telescope sites;
  - Requirements for operational data sharing to mitigate impacts on astronomy; and
  - Consideration of the balance between mission objectives, orbital sustainability, and interference on other space activities.
- The CPS encourages States to take a whole-of-government approach to supporting and encouraging development of mitigations of the impact on astronomy:
  - Support astronomical observatories to develop, test, and operate required mitigation techniques to minimise the impact of large satellite constellations;
  - Implement incentive measures for industry to develop and implement technical and operational techniques to protect astronomy from the negative impacts of large satellite constellations, including predictive tools and test labs to assess brightness using designs or prototype satellites, and basic research on less-reflective alternate materials, as well as the reduction of radio radiation in unwanted directions and at unwanted frequencies.
- The CPS encourages the satellite industry to
  - Continue the collaboration with astronomy to explore the effective co-existence of their services and astronomy;
  - Share mitigation experiments and resources within the satellite industry community to spur the development of affordable, accessible and effective mitigation options;
  - Consider mitigating the impacts on astronomy as an essential component of their long-term sustainability and ESG approaches.

# [ REFERENCES ]

For further details please consult these references:

# [ ACKNOWLEDGEMENTS ]

Gyula I. G. Józsa, MPIfR, Germany
Andrew Williams, ESO / IAU CPS, Intergovernmental Organisation
Richard Green, University of Arizona / IAU CPS, USA
Isabel Marsh, ESO, Intergovernmental Organisation
John Antoniadis, FORTH Institute of Astrophysics, Greece
Domingos Barbosa, ENGAGE SKA, Instituto de Telecomunicações, Portugal
John Barentine, Dark Sky Consulting, LLC / IAU CPS, USA
Guillermo Blanc, Carnegie Institution for Science / Las Campanas Observatory, Chile
Aaron Boley, University of British Columbia / IAU CPS, Canada
Bruno, Coelho, ATLAR Innovation, Portugal
Patricia Cooper, Constellation Advisory, LLC, USA
Sara Dalledonne, ESPI, Austria
Federico Di Vruno, SKAO/IAU CPS, Intergovernmental Organisation
Joe Diamond, SKAO, Intergovernmental Organisation
Adam Dong, University of British Columbia, Canada
Ronald Drimmel, INAF, Italy
Siegfried Eggl, University of Illinois Urbana-Champaign/IAU CPS, USA
Nusrin Habeeb, UAE University, UAE
Jessica Heim, University of Southern Queensland/IAU CPS, Australia
Chris Hofer, Amazon Kuiper, USA
Narae Hwang, KASI, Korea
Mathieu Isidro, SKAO / IAU CPS, Intergovernmental Organisation
David Koplow, Georgetown University, USA
James Lowenthal, Smith College, USA
Sara Lucatello, INAF, Italy
Mariya Lyubenova, ESO, Intergovernmental Organisation
Robert Massey, RAS, UK
Mike Peel, Imperial College London/IAU CPS, UK
Meredith Rawls, University of Washington/Rubin Observatory/IAU CPS, USA
Adrien Saada, Space Sustainability Rating, Switzerland
Alejandro Sanchez, Savestars Consulting S.L.,Spain
Pedro Sanhueza, DECYTI/MINREL, Chile
Warren Skidmore, TMT, USA
Boris Sorokin, SKAO/IAU CPS, Intergovernmental Organisation
P. Sreekumar, Manipal Center for Natural Sciences, India
Tim Stevenson, SKAO/IAU CPS, Intergovernmental Organisation
Paula Tartari, ESO/IAU CPS, Intergovernmental Organisation
Vincenza Tornatore, Politecnico di Milano, Italy
Connie Walker, NSF NOIRLab/IAU CPS, USA
Benjamin Winkel, MPIfR, Germany
Yana Yakushina, Ghent University, Belgium

The lead author’s participation in the development of this document was funded partly from the European Union’s Horizon 2020 research and innovation programme under grant agreement No 101004719.

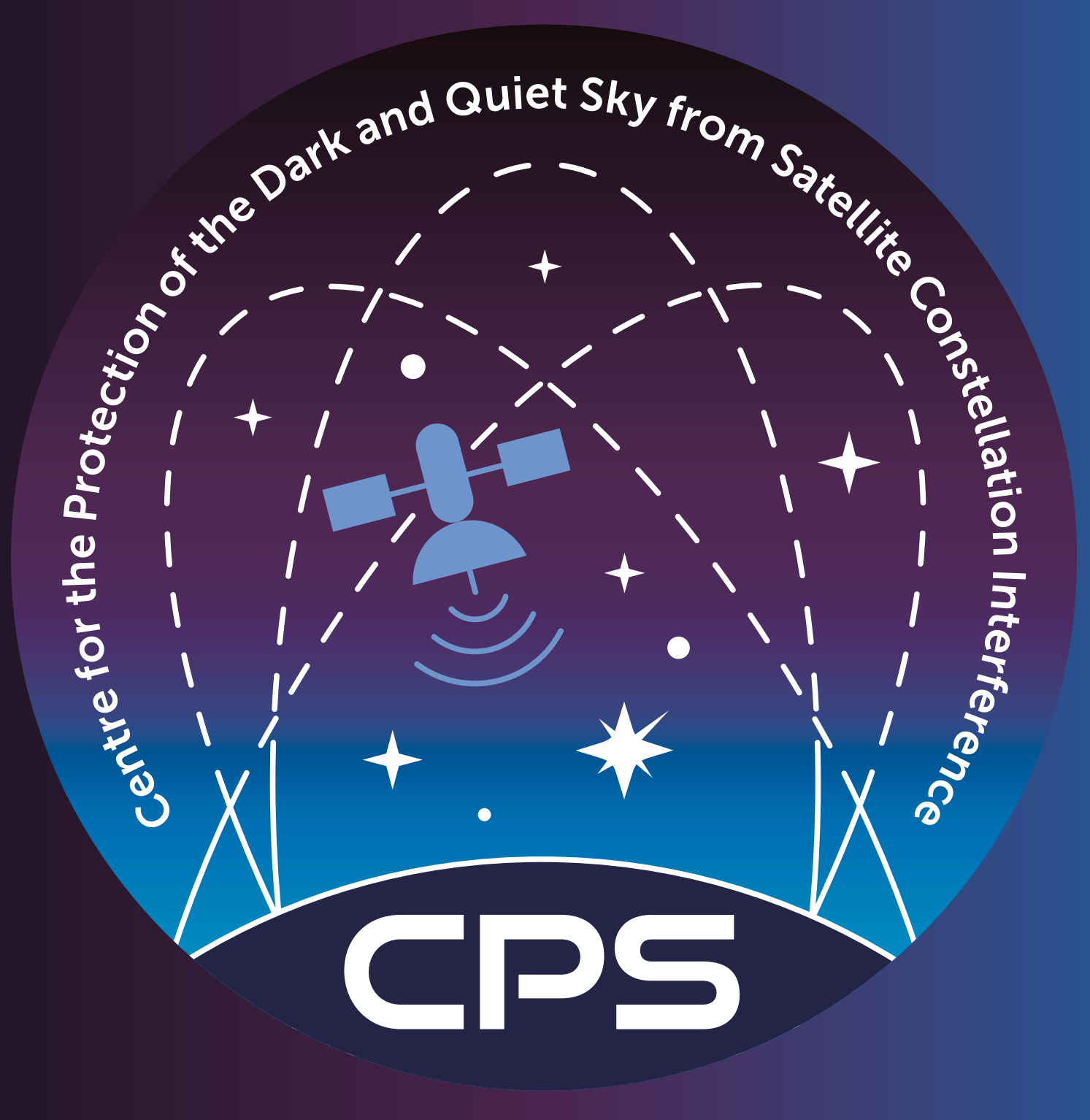

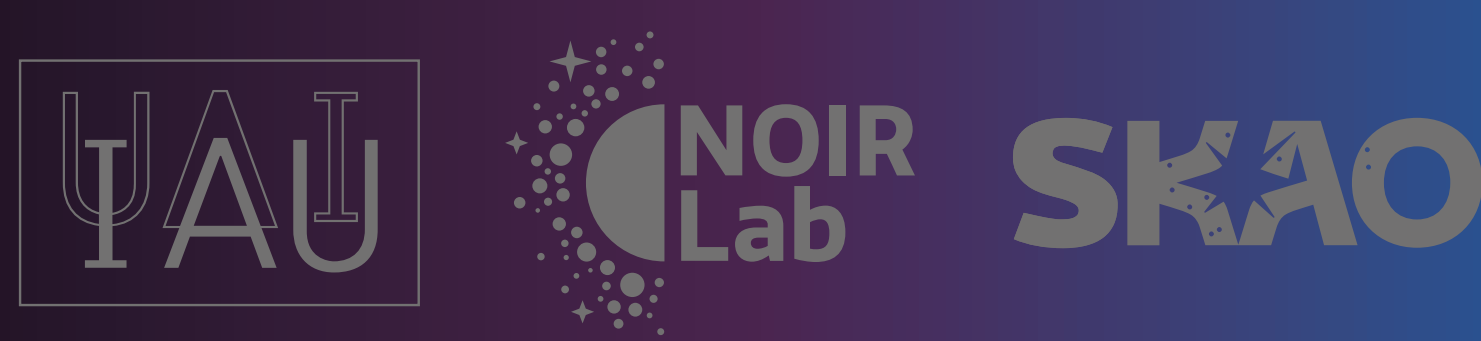

On the front cover: A long-exposure image of the Orion Nebula with a total exposure time of 208 minutes showing satellite trails in mid-December 2019. *Credit: A. H. Abolfath/NOIRLab/NSF/AURA*